\documentclass[twocolumn,showpacs,preprintnumbers,amsmath,amssymb]{revtex4}

\usepackage{graphicx}
\usepackage{dcolumn}
\usepackage{bm}
\usepackage{color}

\begin{document}

\preprint{}

\title{Laser-induced Field Emission from Tungsten Tip: \\Optical Control of Emission Sites and Emission Process}

\author{Hirofumi Yanagisawa${}^{1}$}
\email{hirofumi@physik.uzh.ch}
\author{Christian Hafner${}^{2}$}
\author{Patrick Don\'{a}${}^{1}$}
\author{Martin Kl\"{o}ckner${}^{1}$}
\author{Dominik Leuenberger${}^{1}$}
\author{Thomas Greber${}^{1}$}
\author{J\"{u}rg Osterwalder${}^{1}$}
\author{Matthias Hengsberger${}^{1}$}
\affiliation{${}^1$\mbox{Physik Institut, Universit\"{a}t Z\"{u}rich, Winterthurerstrasse 190, CH-8057 Z\"{u}rich, Switzerland} \\
${}^2$\mbox{Laboratory for Electromagnetic Fields and Microwave Electronics, ETH Z\"{u}rich, Gloriastrasse 35, CH-8092 Z\"{u}rich, Swizerland}}

\begin{abstract}
Field-emission patterns from a clean tungsten tip apex induced by femtosecond laser pulses have been investigated. Strongly asymmetric field-emission intensity distributions are observed depending on three parameters: (1) the polarization of the light, (2) the azimuthal and (3) the polar orientation of the tip apex relative to the laser incidence direction. In effect, we have realized an ultrafast pulsed field-emission source with site selectivity of a few tens of nanometers. Simulations of local fields on the tip apex and of electron emission patterns based on photo-excited nonequilibrium electron distributions explain our observations quantitatively. Electron emission processes are found to depend on laser power and tip voltage. At relatively low laser power and high tip voltage, field-emission after two-photon photo-excitation is the dominant process. At relatively low laser power and low tip voltage, photoemission processes are dominant. As the laser power increases, photoemission from the tip shank becomes noticeable. 
\end{abstract}

\date{\today}
\maketitle

\section{\label{sec:level1}INTRODUCTION}
Field emission from metallic tips with nanometer sharpness has been introduced some time ago as highly bright and coherent electron source \cite{gomer93, fursey03, fink90, fink86, fu01,cho04, oshima02,nagaoka98}. Only recently, pulsed electron sources with high spatio-temporal resolution were realized by laser-induced field emission from such tips \cite{hommelhoff06a,hommelhoff06b, ropers07,barwick07}. Potentially, spatio-temporal resolution down to the single atom and the attosecond range appears to be possible \cite{fink86,fu01,hommelhoff06b}. Such electron sources would be very attractive for applications in time-resolved electron microscopy or scanning probe microscopy. However, the interaction of the laser pulses with the sharp tip and the electron emission mechanism are not yet fully understood \cite{hommelhoff06a, ropers07, barwick07, wu08}.

When a focused laser pulse illuminates the tip, optical electric fields are modified at the tip apex due to the excitation of surface electromagnetic (EM) waves that couple with collective surface charge excitations to form e.g. surface plasmon polaritons. Interference effects of the resulting surface EM waves can lead to \emph{local field enhancement} \cite{aeschlimann07}. Depending on the field strength, different electron emission processes become dominant~\cite{hommelhoff06a}. For relatively weak fields, single electron excitations by single- or multiphoton absorption are dominant, and photo-excited electrons are tunneling through the surface potential barrier; such processes are termed \emph{photo-field emission}. On the other hand, very strong local fields largely modify the tunneling barrier and prompt the field emission directly, leading to \emph{optical field emission}. So far, the different emission processes were disputed in the literature, while the local field enhancement was treated as a static effect such as the \emph{lightening rod effect}~\cite{hommelhoff06a,hommelhoff06b,novotny97, novotny02,hecht05}. Hence, local fields on the tip apex are considered to be symmetric with respect to the tip axis. However, when the tip size is larger than approximately a quarter wavelength, dynamical effects are predicted to occur~\cite{martin01}. 

Here, we used a tip whose apex was approximately a quarter wavelength and we investigated laser-induced field-emission patterns. We have found that dynamical effects substantially influence the symmetries of local field distributions and thereby field-emission intensity distributions. Varying the following three parameters changes these distributions substantially: (1) the laser polarization, (2) the azimuthal and (3) the polar orientation of tip apex relative to the laser incidence direction. These are effects that had not been observed in earlier experiments \cite{gao87, hommelhoff06a}. At the same time, we realized an ultrafast pulsed field-emission source with emission site selectivity on the scale of a few tens of nanometers. In our previous paper, simulations confirm that the photo-field emission process is dominant in laser-induced field emission \cite{yanagisawa09}. Here, we further discuss electron emission processes and their dependence on laser power and tip voltage by investigating electron emission patterns, Fowler-Nordheim plots, and calculated electron energy distributions. At relatively low laser power and high tip voltage, field emission after two-photon photo-excitation is the dominant process. At still relatively low laser power and low tip voltage, multiphoton photoemission over the surface barrier is dominant. As laser power increases, photoemission from the tip shank contributes.

This manuscript consists of four main sections. In section II, we explain our experimental setup and our theoretical model. In section III, we discuss the optical control of field-emission sites and the emission mechanism based on simulations of local fields on the tip apex and laser-induced field-emission microscopy (LFEM) images. In section IV, we discuss the emission processes for varying laser power and tip voltage based on experimental and calculated results. In the last section, we present conclusions and proposals for future experiments.

\section{\label{sec:level1}METHODOLOGY}

\subsection{\label{sec:level1}Experimental setup}

Fig. 1(a) schematically illustrates our experimental setup. A tungsten tip is mounted inside a vacuum chamber ($3 \cdot 10^{-10}$ mbar). Laser pulses are generated in a Ti:sapphire oscillator (center wavelength: 800 nm; repetition rate: 76 MHz; pulse width: 55 fs) and introduced into the vacuum chamber.  An aspherical lens (focal length: $f$ = 18 mm) is mounted on a holder that is movable along the y direction and located just next to the tip to focus the laser onto the tip apex; the diameter of the focused beam is approximately 4 $\mu$m ($1/e^2$ radius) measured by a method with a razor blade~\cite{firester77}. Linearly polarized laser light was used. The polarization vector can be changed within the transversal ($x$, $z$) plane by using a $\lambda/2$ plate~\cite{fowles89}. As shown in the inset, where the laser propagates towards the reader's eye as denoted by the circled dot, the polarization angle $\theta_{P}$ is defined by the angle between the tip axis and the polarization vector.

The tip can be heated to clean the apex and also negatively biased for field emission. Although the tip is polycrystalline tungsten, heating to around 2500$^\circ $C leads to the tip apex being crystallized and oriented towards the (011) direction~\cite{gomer93, sato80}. A position sensitive detector with a Chevron-type double-channelplate amplifier in front of the tip is used to record the emission patterns. The spatial resolution of field emission microscopy (FEM) is approximately 3 nm~\cite{gomer93}. The tip holder can move along three linear axes ($x$, $y$, $z$) and has two rotational axes for azimuthal ($\varphi$, around the tip axis) and polar ($\theta$, around the $z$ axis) angles~\cite{greber97}. The laser propagates parallel to the horizontal $y$ axis within an error of $\pm$ 1 degree. $\theta$ is set so that the tip axis is orthogonal to the laser propagation axis. The orthogonal angle in $\theta$ was determined by plotting positions of the tip in ($x$, $y$) coordinates while keeping the tip apex in the focus of the laser as schematically shown in Fig. 1(b). The maximum position in $x$ gives the orthogonal angle. Experimental data are shown in Fig. 1(c). The plots were taken in 0.5 degree steps. The data clearly show the maximum in the $x$ position. We defined the corresponding angle as $\theta = 0$ for convenience. The precision is estimated to be $\pm$ 1 degree. In these experiments, the base line of the rectangular detector is approximately 20$^\circ$ off from the horizontal ($y$ axis) incidence direction, which means that the laser propagation axis is inclined by 20$^\circ$ from the horizontal line in the observed laser-induced FEM images (see dashed red arrow in Fig. 3a). All the measurements were done at room temperature.

\begin{figure}[t!]
\begin{center}
\includegraphics[scale=0.21]{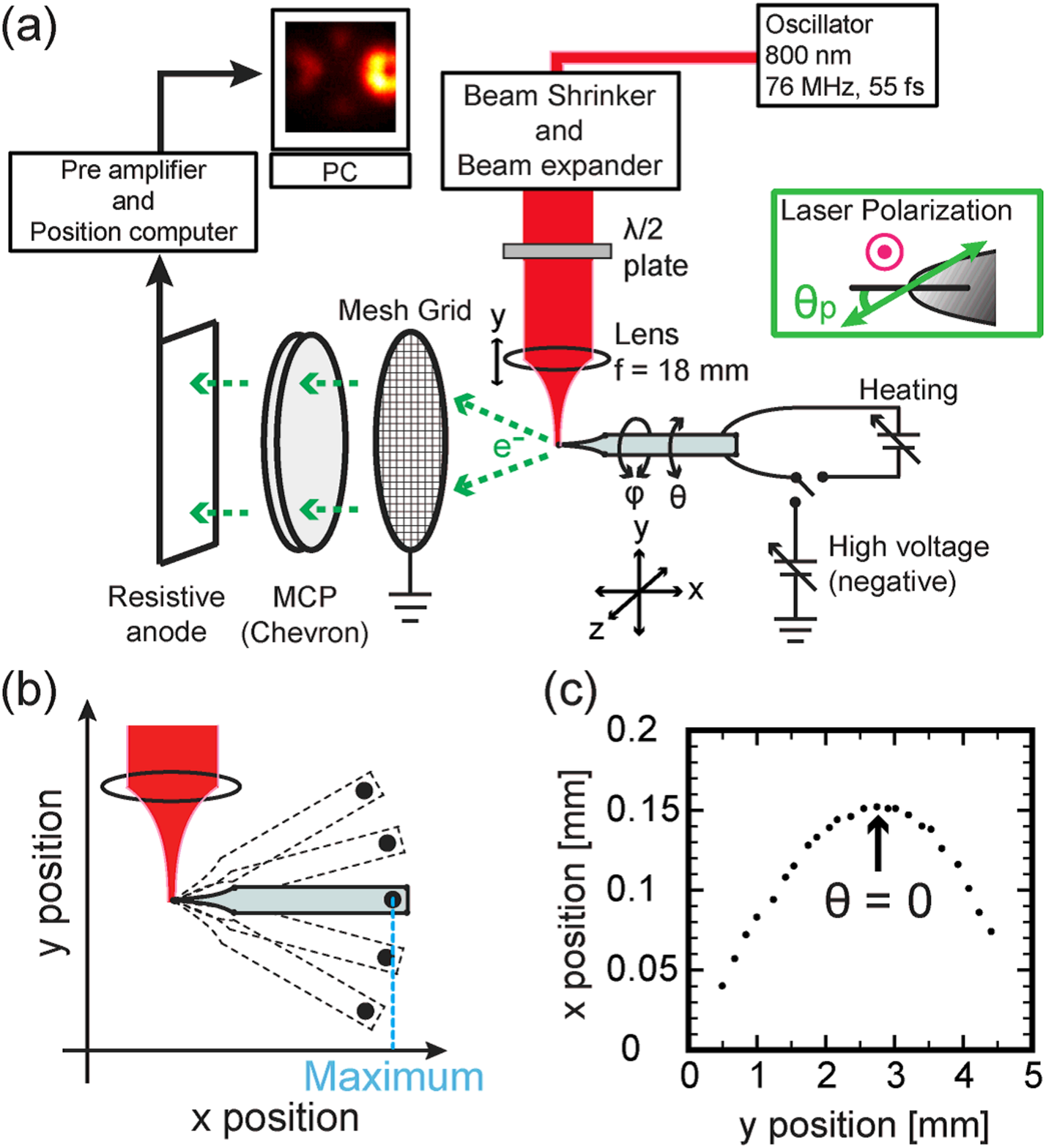}
\end{center}
\vskip -\lastskip \vskip -3pt
\caption{\label{fig:epsart}
(color online). Schematic diagram of the experimental setup (a). A tungsten tip is mounted inside a vacuum chamber. Laser pulses are generated outside the vacuum chamber. An aspherical lens is located just next to the tip to focus the laser onto the tip apex. Emitted electrons are detected by a position-sensitive detector in front of the tip. The polarization angle $\theta_{P}$ is defined in the inset, where the laser beam propagates towards the reader's eye (see text for further description). (b) shows a schematic diagram defining the orthogonal angle between the laser propagation direction and the tip axis. The right angle is where x is maximum. (c) shows tip positions in (x, y) coordinates for different $\theta$, found while the tip apex is kept in the focus of the laser. The angle which gives maximum x is defined as $\theta = 0$ for convenience.}
\label{fig:label-1}
\end{figure}

\subsection{\label{sec:level1}Theoretical model}

\begin{figure}[b]
\begin{center}
\includegraphics[scale=0.26]{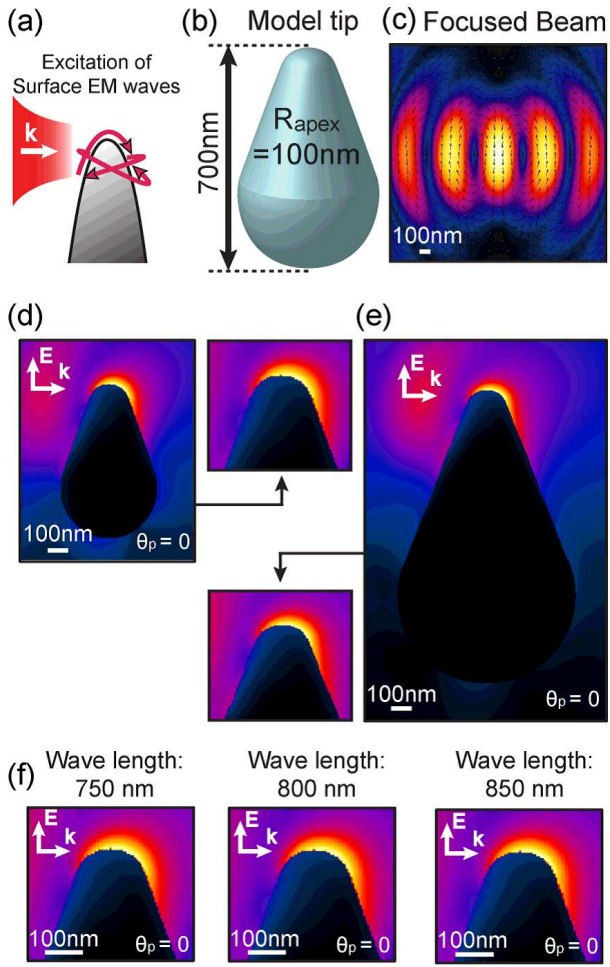}
\end{center}
\vskip -\lastskip \vskip -3pt
\caption{\label{fig:epsart}
(color online). Schematic illustration of the excitation and interference of surface EM waves (a) and of the model tip (b). The radius of curvature of the tip apex is 100 nm, and the length is 700 nm. (c) represents the field distribution of the focused beam used in the simulation at a certain time. The beam waist is 1 $\mu$m and the wavelength is 800 nm. Small arrows indicate the field direction, and field strength is represented by a linear color scale: highest field values are represented in yellow (brightest color). The calculated time-averaged field distribution around the model tip is shown in (e) for $\theta_{P} = 0^\circ$ together with a magnified picture in the vicinity of the tip apex. (e) shows the time-averaged field distribution around a longer model tip together with a magnified picture in the vicinity of the tip apex; the tip length is twice as large as that of (d). (f) represents the time-averaged field distributions around the model tip simulated by incident laser light with wavelengths of 750 nm, 800 nm and 850 nm. }
\label{fig:label-5}
\end{figure}

Although the field emission is a quantum mechanical phenomenon, the interaction between the optical fields and the tungsten tip apex can be treated classically by solving Maxwell equations. Such an interaction can be understood by a mechanistic picture as shown in Fig. 2(a). When a laser pulse illuminates the metallic tip, surface EM waves are excited, which propagate around the tip apex. As a result of the interference among the excited waves, the optical fields are modulated. To simulate the superposition of surface EM waves and the resulting local field distributions on the tip apex, we used the Multiple Multipole Program (MMP)~\cite{christian90,christian99,mmp}, which is a highly accurate semi-analytic Maxwell solver, available in the package MaX-1~\cite{max1} and in the open source project OpenMaX~\cite{christian98,openmax1}.

A droplet-like shape was employed as a model tip as shown in Fig. 2(b), with a radius of curvature of the tip apex of 100 nm, which is a typical value for a clean tungsten tip. Atomic structures were not included in the model becasue the tip apex can be regarded as a smooth surface on this length scale given by the tip dimensions and the wavelength of the laser field. The dielectric function $\epsilon$ of tungsten at 800 nm was used, i.e. a real part $Re(\epsilon) = 5.2$ and an imaginary part $Im(\epsilon) = 19.4$ \cite{tungstenepsilon}. Note that accuracy of the dielectric functions does not affect our conclusion, which will be demonstrated in section III B by comparing with local fields on a gold tip. A focused laser with a beam waist of 1 $\mu$m and a wavelength of 800 nm was used as shown in Fig. 2(c). The model tip was set so that its apex is at the center of the focus.

By using different droplet sizes it was verified that the model tip is long enough so as to mimick the infinite length of the real tip; the fields at the truncated side of the tip are substantially weaker so that the excited surface EM waves propagating arond the whole tip do not affect the induced field distribution at the tip apex. Figure 2(d) shows the calculated time-averaged field distribution around the model tip of Fig. 2(b). Fig. 2(e) shows the same calculated field distribution for a longer tip with the same radius of curvature of the tip apex of 100 nm. In both cases, the laser is propagating from left to right, where the polarization vector has been chosen parallel to the tip axis ($\theta_{P} = 0^\circ$). The magnified pictures around the tip apex of Fig. 2(d) and 2(e) show that the local field distributions of the two are basically the same, indicating the length of the model tip in Fig. 2(b) to be long enough.

In the simulations, we only used a center wavelength of 800 nm, even though the laser pulse has a spectral width of $\Delta = 25$ nm with respect to the center wavelength. Justification of the use of only a center wavelength is done by simulating the local electric fields with laser light of wavelengths of 750 nm and 850 nm. Fig. 2(f) shows the time-averaged field distributions around the model tip obtained by excitation with these three wavelengths. They are almost the same, which indicates that the substantial spectral width of the light around 800 nm would not affect the position of the maximum local electric fields simulated with the wavelength of 800 nm.

\section{\label{sec:level1}OPTICAL CONTROL OF FIELD-EMISSION SITES and EMISSION MECHANISM}

\subsection{\label{sec:level1}Experimental Results}

\begin{figure}[b!]
\begin{center}
\includegraphics[scale=0.31]{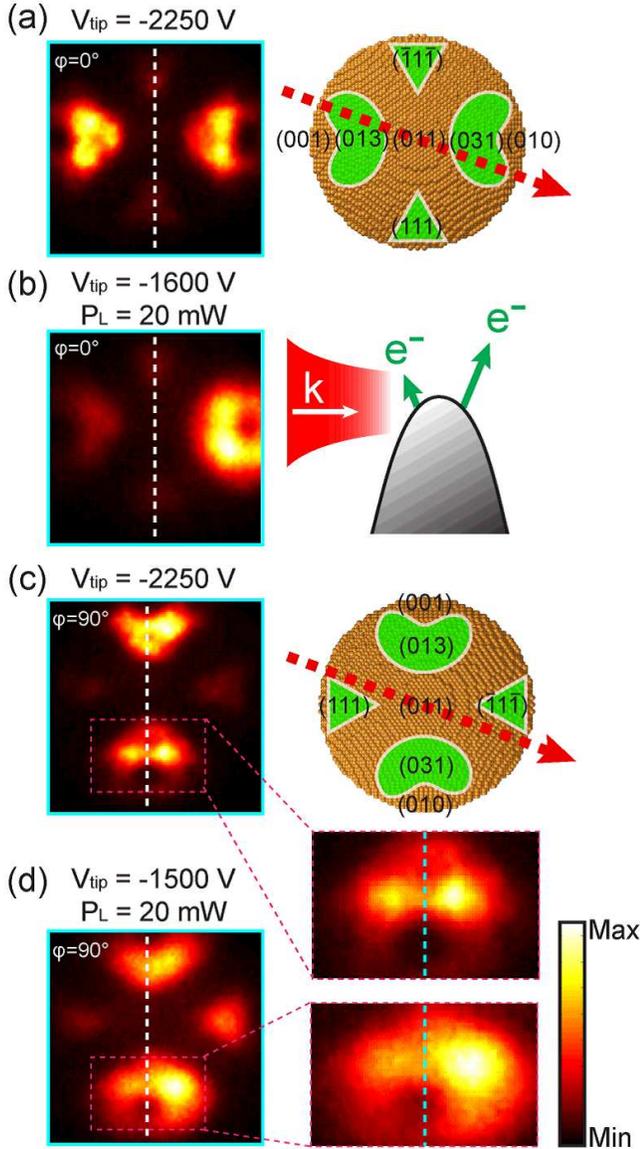}
\end{center}
\vskip -\lastskip \vskip -3pt
\caption{\label{fig:epsart}
(color online). Electron emission patterns for two orthogonal azimuthal orientations ($\varphi$) of the tip without laser [$\varphi = 0^\circ$ (a) and $\varphi = 90^\circ$ (c)], and with laser irradiation [$\varphi = 0^\circ$ (b) and $\varphi = 90^\circ$ (d)]. $V_{tip}$ indicates the DC potential applied to the tip and $P_{L}$ indicates the laser power measured outside the vacuum chamber. The insets of (a) and (c) show the front view of the atomic structure of a tip apex with a curvature radius of 100 nm, based on a ball model, in which green areas with white edges indicate the field emission sites and the red arrow indicates the laser propagation direction. The inset of (b) shows a schematic side view of the laser-induced field emission geometry, in which green vectors indicate intensities of electron emission and the white arrow indicates the laser propagation direction. A dashed white line denotes a mirror symmetry line of the atomic structure in each picture. In (c) and (d) specific regions of interest, marked by dashed red rectangles, are blown up on the right hand side. }
\label{fig:label-2}
\end{figure}

  The field emission pattern from the clean tungsten tip apex which orients towards the (011) direction is shown in Fig. 3(a). The most intense electron emission is observed around the (310)-type facets, and weaker emission from (111)-type facets. These regions are highlighted by green areas with white edges on the schematic front view of the tip apex in the inset of Fig. 3(a). The intensity map roughly represents a work function map of the tip apex: the lower the work function is, the more electrons are emitted. The relatively high work functions of (011)- and (001)-type facets \cite{michaelson77} suppress the field emission from those regions.

The laser-induced FEM (LFEM) image in Fig. 3(b), taken with the light polarization oriented parallel to the tip axis ($\theta_{P}$ = 0), shows a striking difference in symmetry compared to that of the FEM image in Fig. 3(a). Emission sites are the same in both cases, but the emission pattern becomes strongly asymmetric with respect to the shadow (right) and exposed (left) sides to the laser incidence direction. The most intense emission is observed on the shadow side as illustrated in the inset of Fig. 3(b). Figs. 3(c) and 3(d) give the same comparison for a different azimuthal orientation of the tip as shown in the inset of Fig. 3(c). In the magnified image of Fig. 3(c), two emission sites can be identified which are separated by approximately 30 nm. The strong emission asymmetry is observed even over such short distances, as shown in Fig. 3(d). Actually, the laser pulses arrive at an angle of 20$^\circ$ off the horizontal line in both LFEM images, as indicated by the dashed red arrow in the inset of Fig. 3(a). This oblique incidence slightly affects the symmetry with respect to the central horizontal line in the observed LFEM images (see below).

\begin{figure}[b]
\begin{center}
\includegraphics[scale=0.22]{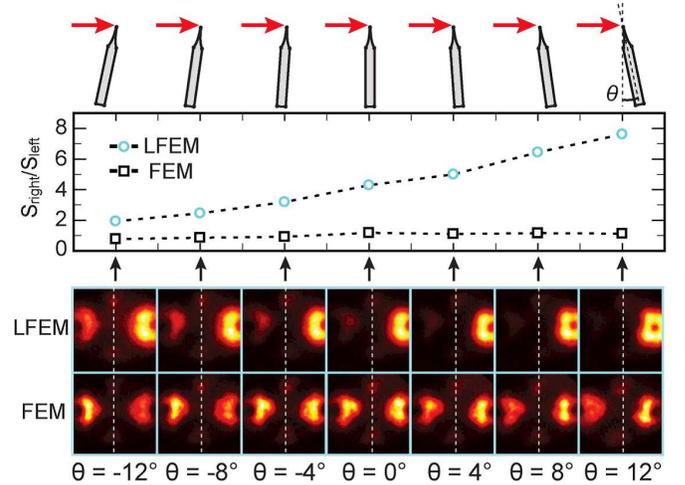}
\end{center}
\vskip -\lastskip \vskip -3pt
\caption{\label{fig:epsart}
(color online). $\theta$-dependence of LFEM images at [$\varphi = 0^\circ$, $\theta_{P} = 0^\circ$], which were taken at $V_{tip} =$ -1500 V and $P_{L} =$ 20 mW. $\theta$ is varied from $\theta = -12^\circ$ to $\theta = 12^\circ$ by $4^{\circ}$ steps. Schematic diagrams for the experimental configuration are shown at the top, in which red arrows indicate the laser propagation direction. The corresponding FEM images are also shown below, which were taken at $V_{tip} =$ -2200 V. The white dashed lines in the pictures denotes a mirror symmetry line of the atomic structure. The total yield $S_{right}$ from right side of each image and the total yield from left side $S_{left}$  with respect to the white dashed line were taken. The ratio of $S_{right}$ to $S_{left}$ are plotted in the graph. Blue circles are for LFEM and black squres are for FEM.}
\label{fig:label-4}
\end{figure}


\begin{figure*}[t!]
\begin{center}
\includegraphics[scale=0.30]{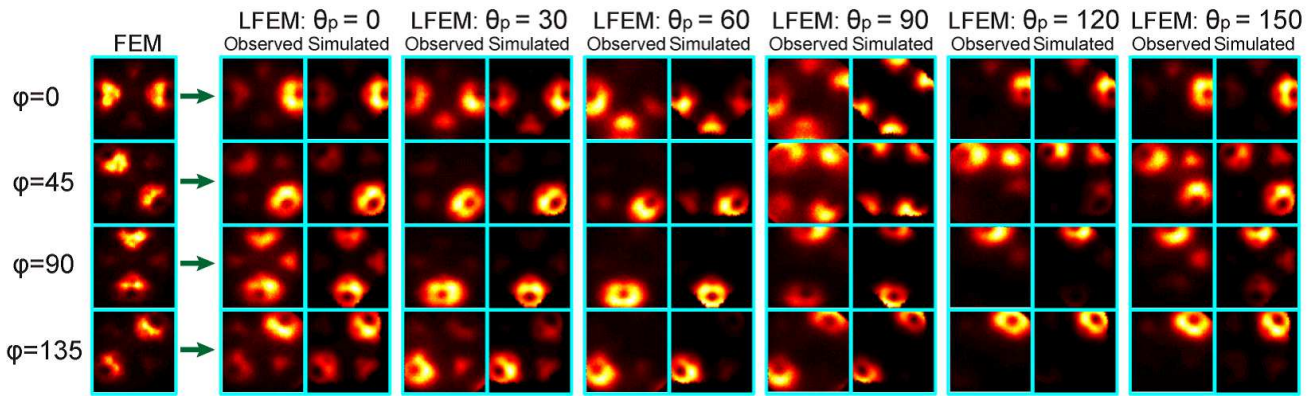}
\end{center}
\vskip -\lastskip \vskip -3pt
\caption{\label{fig:epsart}
(color online). Comparison of measured and simulated laser-induced FEM (LFEM) images for different light polarization angles $\theta_{P}$ and for different azimuthal orientations $\varphi$ of the tip. The leftmost column gives the FEM images without laser irradiation for four different azimuthal angles ($V_{tip} =$ -2250 V). For the same azimuthal angles, observed LFEM  images are shown as a function of polarization angle $\theta_{P}$ in 30$^\circ$ steps ($V_{tip} \approx$ -1500 V and $P_{L} =$ 20 mW). The simulated LFEM images from the photo-field emission model, in which $V_{tip}$ and $P_{L}$ were set as in the corresponding experiments, are shown on the right-hand side of the observed LFEM images. The color scale and laser propagation direction are the same as in Fig. 3.}
\label{fig:label-3}
\end{figure*}

The asymmetry in LFEM images can be controlled further by changing $\theta$. In Fig. 4, the $\theta$-dependence of LFEM images at [$\varphi = 0^\circ$, $\theta_{P} = 0^\circ$] is shown, which were taken at $V_{tip} = $ -1500 V and $P_{L} = $ 20 mW. $\theta$ is varied from $\theta = -12^\circ$ to $\theta = 12^\circ$ by 4 degree steps. Schematic diagrams for the experimental configuration are shown at the top, in which red arrows indicate the laser propagation direction. The corresponding FEM images, which were taken at $V_{tip} = $ -2200 V, are also shown. As $\theta$ increases, the asymmetry of the LFEM images becomes clearly enhanced. At $\theta = 12^\circ$, electrons are emitted almost only from right-side emission sites. Among these $\theta$ values, the symmetry of the FEM images changed only slightly due to a change of the DC field distribution on the tip apex. To distinguish between the contributions of DC and laser field distributions to the asymmetry of the LFEM images, we evaluated the change in symmetry quantitatively. The total yield $S_{right}$ from the right side of each image and the total yield from the left side $S_{left}$ were obtained from each image with respect to the white dashed line. Then the ratios of $S_{right}$ to $S_{left}$ are plotted in the graph: high values indicate large asymmetries. The asymmetry is clearly enhanced in LFEM with respect to FEM, which indicates that the laser fields mainly contribute to enhance the asymmetry for higher angles $\theta$.

We also found experimentally a strong dependence of the electron emission patterns on the laser polarization direction and on the azimuthal tip orientation. Fig. 4 shows the LFEM patterns for different values of $\theta_{P}$ in 30$^\circ$ steps, and for four different azimuthal orientations $\varphi$ of the tip. The corresponding FEM images are also shown in the left-most column; they show simply the azimuthal rotation of the low work function facets around the tip axis. Throughout the whole image series, the emission sites do not change, but intensities are strongly modulated resulting in highly asymmetric features. For instance, for [$\varphi = 0^\circ$, $\theta_{P} = 0^\circ$] the intense emission sites are located on the right-hand (shadow) side of the tip, but for [$\varphi = 0^\circ$, $\theta_{P} = 60^\circ$] the emission sites on the left-hand side become dominant. LFEM images recorded for $\theta_{P} = 1 80$ (not shown) are exactly the same as those for $\theta_{P} = 0^\circ$, and all the LFEM images are well reproducible.


\subsection{\label{sec:level1}Simulations of Local fields}

\begin{figure}[th!]
\begin{center}
\includegraphics[scale=0.23]{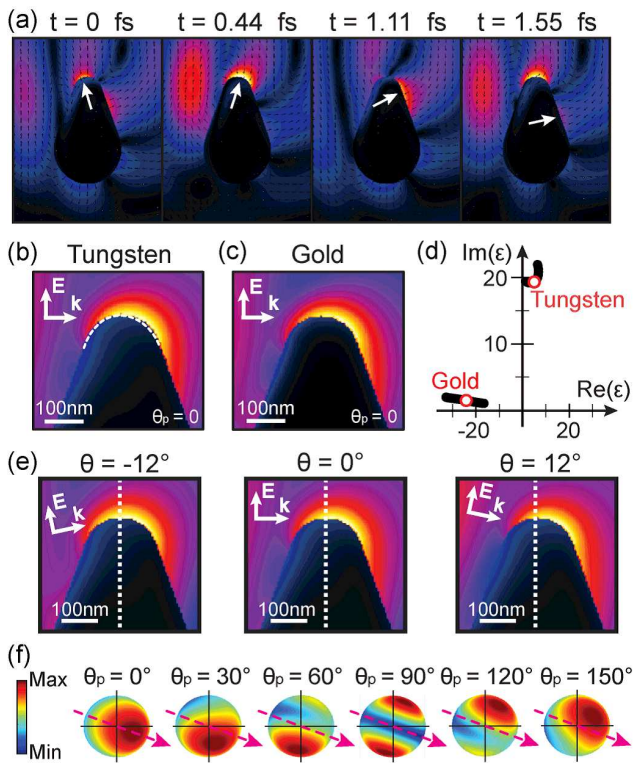}
\end{center}
\vskip -\lastskip \vskip -3pt
\caption{\label{fig:epsart}
(color online). Time evolution of laser fields over a cross section of the model tip while propagating through the tip apex from left to right (a). The polarization vector is parallel to the tip axis ($\theta_{P} = 0^\circ$). Small black arrows indicate the field direction, and field strength is represented by a linear color scale: highest field values are represented in yellow (brightest color). The time-averaged field distribution for tungsten and gold tips are shown in (b) and (c) where the model tip of Fig. 2(b) is employed for both. The dielectric function of tungsten and gold for the wavelengths between 700 nm and 900 nm are plotted by black dots in (d) and the values at 800 nm are highlighted by red circles. (e) shows the time-averaged field distributions around the tungsten tip for three different polar angles: $\theta = -12^\circ$, $\theta = 0^\circ$, and $\theta = 12^\circ$. In (f) the time-averaged field distributions are given in a front view of the model tip for different polarization directions $\theta_{P}$ ($\theta = 0^\circ$). The laser propagation direction is indicated by red arrows, and is the same as in the experiment.}
\label{fig:label-5}
\end{figure}

When a laser pulse illuminates the metallic tip, surface EM waves are excited, which propagate around the tip apex as illustrated schematically in Fig. 2(a). Due to the resulting interference pattern, the electric fields show an asymmetric distribution over the tip apex, depending also on the laser polarization. Fig. 6(a) shows the time evolution of laser fields at 800 nm wavelength over a cross section of the model tip while propagating through the tip apex from left to right, where the polarization vector has been chosen parallel to the tip axis ($\theta_{P} = 0^\circ$). It can be seen that a surface EM wave is propagating around the tip apex indicated by white arrows and enhanced at the tip apex. The calculated time-averaged field distribution around the tip apex is shown in Fig. 6(b). The field distribution is clearly asymmetric with respect to the tip axis, with a maximum on the shadow side of the tip. The field enhancement factor of the maximum point is 2.5 with respect to the maximum field value of the incident laser, and 1.7 for the counterpart of the maximum point on the side exposed to the laser. This is consistent with our observations in Fig. 3(b) and 3(d) where the field emission is enhanced on the shadow side. 

Additionally, we would like to note that a similar asymmetric distribution can also be seen even for a metal with a dielectric function which is largely different from that of tungsten. For example, we performed a simulation for gold using a real part $Re(\epsilon) = -24$ and an imaginary part $Im(\epsilon) = 1.5$ \cite{goldepsilon}. Fig. 6(c) shows the time-averaged field distribution on the gold tip apex. The field distribution shows a similar asymmetry as for tungsten. It should also be mentioned that surface EM waves are classified in terms of the dielectric functions of the interacting material~\cite{zenneck} though some authours do not distinguish. If the real part of the dielectric function is negative, then the surface EM waves are proper \emph{surface plasmon polaritons}. On the other hand, if the real part is positive and the imaginary part is large, the term \emph{Zenneck waves} is more appropriate. The dielectric functions of tungsten and gold between 700 nm and 900 nm are plotted by black dots in Fig. 6(d), where the values at 800 nm are highlighted by red circles. From Fig. 6(d), strictly speaking, the excited surface EM waves on tungsten are Zenneck waves and those on gold are surface plasmon polaritions. Figs. 6(b) and (c) also indicate that different kinds of surface EM waves do not show substantial difference in the resulting field distribution.

The asymmetric local field distribution can be controlled by changing the polar angle $\theta$ and the laser polarization angle $\theta_{P}$. Fig. 6(e) shows time-averaged field distributions on the tungsten tip apex with different laser incidence directions relative to the polar orientation of the tip apex. As $\theta$ increases, the asymmetry becomes stronger. This is consistent with our observation in Fig. 4 where the most asymmetric emission is observed at $\theta = 12^\circ$. Fig. 6(f) shows, in a front view, time-averaged field distribution maps from the white dashed line region of the model tip in Fig. 6(b). This area corresponds roughly to the observed area in our experiments. The red arrows indicate the laser propagation direction which has been set to the same as in our experimental situation. The field distribution changes strongly depending on the polarization angle. While the maximum field is located directly on the shadow side of the tip for $\theta_{P} = 0^\circ$, the maximum moves towards the lower side of the graphs in concert with the polarization vector for $\theta_{P} = 30^\circ$ and $\theta_{P} =60^\circ$, and reappears on the upper side for $\theta_{P} = 120^\circ$ and $\theta_{P} = 150^\circ$. For $\theta_{P} = 90^\circ$ the polarization vector is perpendicular to the tip axis and produces two symmetric field lobes away from the tip apex. In general the observed LFEM images show the same intensity modulations (Fig. 5): each LFEM pattern at $\theta_{P} = 30^\circ$ and $60^\circ$ shows pronounced emission at the lower side of the image, while each LFEM pattern at $\theta_{P} = 120^\circ$ and $150^\circ$ has maximum emission at the upper side of the image.

\subsection{\label{sec:level1}Simulations of LFEM by the photo-field emission model}

From the calculated local fields, we further simulated the LFEM images by considering the photo-field emission mechanism. The current density $j_{calc}$ of field emission can be described in the Fowler-Nordheim theory based on the free-electron model as follows~\cite{gomer93,murphy56,young59,fursey03},
\begin{eqnarray}
j_{calc} =  \frac{em}{2 \pi^{2} \hbar^{3}} \int_{-W_a}^{\infty} \int_{-W_a}^{W = E} D(W, \Phi, F)f(E)\,d W d E.
\end{eqnarray}
Here, $e$ is the electron charge and $m$ the electron mass, $-W_{a}$ is the effective constant potential energy inside the metal, W is the normal energy with respect to the surface, and E is the total energy. Important factors are $D(W, \Phi, F)$ and $f(E)$. $D(W, \Phi, F)$ is the probability that an electron with the normal energy W penetrates the surface barrier. It depends exponentially on the triangular-shaped potential barrier above W, as indicated by the cross-hatched area in Fig. 7(a) where field emission with energy W occurs. The cross-hatched area is determined by the work function $\Phi$ and the electric field $F$ just outside the surface. $f(E)$ is an electron distribution function. In the case of field emission we have $F = F_{DC}$, where $F_{DC}$ is the applied DC electric field, and $f(E)$ is the Fermi-Dirac distribution at 300 K as shown in Fig. 7(a).  
\begin{figure}[t]
\begin{center}
\includegraphics[scale=0.22]{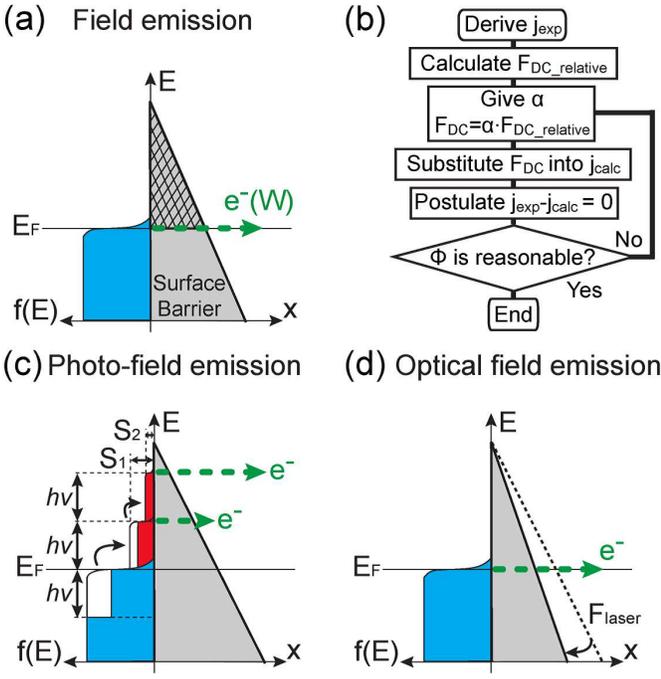}
\end{center}
\vskip -\lastskip \vskip -3pt
\caption{\label{fig:epsart}
(color online). A schematic diagram of field emission from a Fermi-Dirac distribution (a), where an electron with a normal energy W is emitted. The surface barrier above W is shown by a cross-hatched area. (b) shows the logic diagram for obtaining work function and absolute DC field maps. (c) and (d) show schematic diagrams of photo-field emission from a nonequilibrium electron distribution and optical field emission from a Fermi-Dirac distribution, respectively.}
\label{fig:label-5}
\end{figure}
In the photo-field emission model, $F$ still equals $F_{DC}$, but the electron distribution is strongly modified by the electron-hole pair excitations due to single- and multi-photon absorption, resulting in a nonequilibrium distribution characterized by a steplike profile, as illustrated in Fig. 7(c)~\cite{wu08, lisowski04}. For example, one-photon absorption creates a step of height $S_{1}$ from $E_{F}$ to $E_{F}+h\nu$ by exciting electrons from occupied states between $E_{F}-h\nu$ and $E_{F}$. Absorption of a second photon creates a step of height $S_{2}$ from $E_{F}+h\nu$ to $E_{F}+2h\nu$, where $S_{2} \approx S_{1}^2$. We included absorption of up to four photons. The step height $S_{1}$ is proportional to the light intensity $I$. In the vicinity of the tip we have $I\propto F_{laser}^2$ where $F_{laser}$ is the enhanced optical electric field that varies over the tip apex as illustrated in Fig. 6(f) \cite{merschdorf04}.

\begin{figure}[t]
\begin{center}
\includegraphics[scale=0.19]{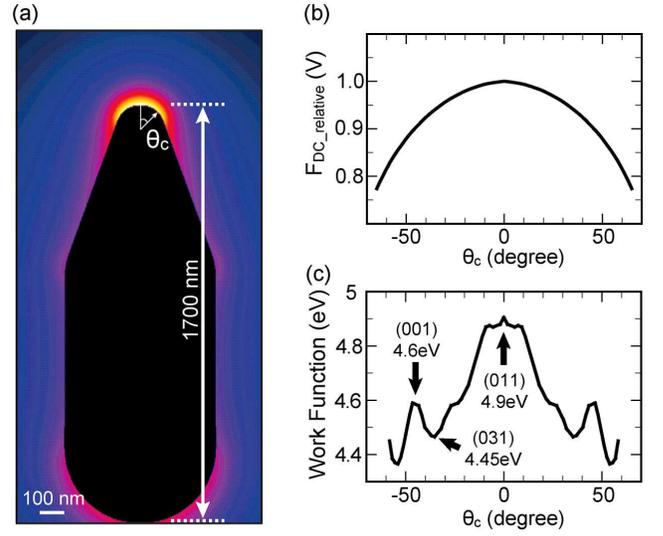}
\end{center}
\vskip -\lastskip \vskip -3pt
\caption{\label{fig:epsart}
(color online). The calculated DC field distribution around the model tip is shown in (a). A wire-like shape was employed for the simulation of relative DC fields. The color scale is the same as in Fig. 6. The relative DC field distribution at the tip apex of (a) is shown as a function of angle $\theta_{c}$, which is defined in (a). The obtained work function profile along a (001)-(011)-(010) curve is shown in (c) as a function of $\theta_{c}$.}
\label{fig:label-5}
\end{figure}

\begin{table}[b]
\caption{\label{tab:table1}Table of work functions of tungsten for several faces. An error in experimental values can be as much as $\pm$ 0.3 eV \cite{hufner95}. }
\begin{ruledtabular}
\begin{tabular}{lcccc}
Facet type &(011) &(001)& (031)  \\
\hline
Simulation &4.9 eV & 4.6 eV & 4.45 eV \\
Experiment &5.25 eV\footnotemark[1] & 4.63 eV\footnotemark[1] & 4.35 eV\footnotemark[2]\\
\end{tabular}
\end{ruledtabular}
\footnotetext[1]{Reference~\cite{michaelson77}.}
\footnotetext[2]{Reference~\cite{Mendenhall34}.}
\footnotetext[3]{Reference~\cite{hufner95}.}
\end{table}

There are three adjustable parameters in our calculations of $j_{calc}$: $\Phi$, $F_{DC}$, and $S_{1}$. They are all functions of position on the tip apex. $\Phi$ and $F_{DC}$ maps on the tip apex were obtained from the measured FEM images by following the logic diagram shown in Fig. 7(b). The measured FEM images, which were symmetrized to have the ideal two-fold symmetry, represent the current density $j_{exp}$ as a function of position on the tip apex, because the electrons follow closely the field lines from the tip apex to the position sensitive detector. In practice we assumed a radius of curvature of the tip apex of 100 nm and used the FEM image at $\varphi = 45^\circ$ shown in Fig. 5. Second, a relative DC field $F_{DC\_relative}$ distribution was generated by MaX-1. We used a more wire-like tip shape for this purpose as shown in Fig. 8(a), and a grounded plate was set 1 cm away from the tip, which is close to that in our experimental setup. The simulated $F_{DC\_relative}$ is shown in Fig. 8(b), which is normalized by the value at tip apex. Going away from the tip apex $F_{DC\_relative}$ decreases. A scaling factor $\alpha$ is introduced, which determines $F_{DC}$ by $F_{DC} = \alpha \cdot F_{DC \_ relative}$. We then obtained the $\Phi$ map by inserting $F_{DC}$ into Eq. (1) and postulating $j_{exp} -  j_{calc} = 0$. The resulting $\Phi$ map was compared to known values for several surface facets of tungsten. The scaling factor $\alpha$ was changed and the procedure was iterated until reasonable work functions were obtained. Thus, a full $\Phi$ map and absolute values for $F_{DC}$ were determined. 

A line profile of the resulting $\Phi$ map along the (001)-(011)-(010) curve is shown in Fig. 8(c). The work function has local curve maxima at the (011)- and (001)- type facets and local minima at (310)- type facets, which is in line with the observed field emission intensity distribution seen in Fig. 3(a); the higher the work function, the lower the intensity. The resulting $\Phi$ values are summarized for several facets and compared with known experimental values in Table 1. They are in fair agreement with each other. A field strength $F_{DC}$ of $2.15$ V/nm results at the tip apex center for the FEM image taken with $V_{tip} = -2250$ V, which is a typical value for FEM. The LFEM experiments were carried out with a reduced tip voltage $V_{tip} \approx -1500$ V. Therefore we used a down-scaled value of 1.43 V/nm in the LFEM simulations. Note that the uncertainty in the $\Phi$ values is not important for our conclusions which will be discussed below: we have also checked the whole discussion in this section with a different work function map using 4.6 eV, 4.32 eV, 4.20 eV for (011), (001) and (310) facets, respectively, but the main outcome does not change.

Substituting the obtained $\Phi$ and $F$ distribution maps into Eq. (1), and using a nonequilibrium electron distribution $f(E)$, the absolute values of $S_{1}$ over the tip apex were determined by fitting the measured total current from the (310) facet on the right-hand side of the LFEM image in Fig. 3(b). The resulting maximum value for $S_{1}$ was $1.6 \cdot 10^{-6}$. By substituting all the adjusted parameters into Eq. (1), we could simulate all the LFEM images. The calculated current densities on the tip apex were projected to the flat screen by following the static field lines. The simulated images can now directly be compared to the experimental images (Fig. 5): they are in excellent agreement in every detail. This comparison clearly demonstrates that the observed strongly asymmetric features originate from the modulation of the local photo-fields.

\subsection{\label{sec:level1}Simulations of LFEM by the optical field emission model}

We also simulated the LFEM images for the optical field emission process and compared the resulting intensity distributions to those of the photo-field emission model. In the optical field emission model schematically shown in Fig. 7(d), the Fermi-Dirac distribution is not modified, but instead the electric field $F$ in Eq. (1) is expressed as $F = F_{DC}+F_{laser}^{\perp}$ where $F_{laser}^{\perp}$ is the normal component of $F_{laser}$ at each point of the tip surface. The absolute values for $F_{laser}^{\perp}$ on the tip apex were determined in the same way as described above for $S_{1}$. The resulting maximum value for $F_{laser}^{\perp}$ was 0.71 V/nm. 

Fig. 9(a) shows the LFEM image for the optical field emission process together with experimentally obtained LFEM and the simulated LFEM image based on the photo-field emission model at [$\varphi = 0^\circ$, $\theta_{P} = 0^\circ$]. The optical field emission model results in an even more strongly asymmetric pattern as compared to the photo-field emission model. This contrasts with the experimental data. Fig. 9(b) shows line profiles extracted from the observed FEM and LFEM images, and from the corresponding simulations for both models, which are all normalized to their maximum value. The measured LFEM profile clearly shows the asymmetric feature observed in the regions B and D. The photo-field emission model catches this asymmetry much more quantitatively than the optical field emission model, as can be best seen in region D. Therefore, the local fields in our experiment are still weak enough such that the photo-field emission process is the dominant one.

\begin{figure}[t]
\begin{center}
\includegraphics[scale=0.18]{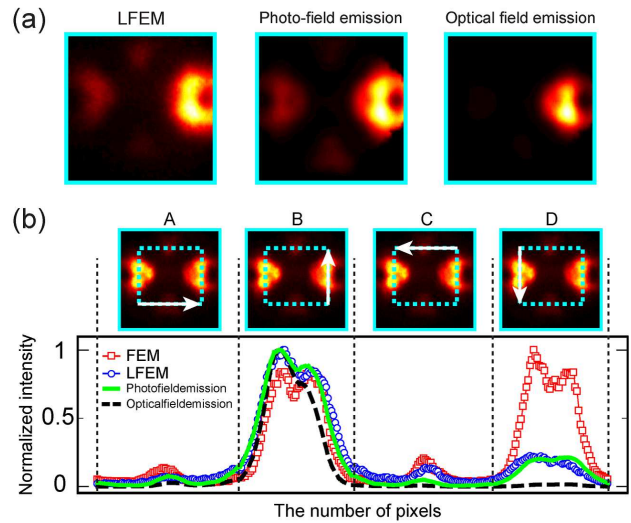}
\end{center}
\vskip -\lastskip \vskip -3pt
\caption{\label{fig:epsart}
(color online). (a) experimentally obtained LFEM image and simulated LFEM image at [$\varphi = 0^\circ$, $\theta_{P} = 0^\circ$] for both photo-field emission and optical field emission models. (b) shows line profiles extracted from the observed FEM images (red line with squares), LFEM images (blue line with circles), and from LFEM images simulated by the photo-field emission model (green solid line) and the optical field emission model (black dashed line) at [$\varphi = 0^\circ$, $\theta_{P} = 0^\circ$]. The whole scanned line corresponds to the unfolded rectangle indicated by the dashed blue line in the FEM figures above, and the corresponding sides are indicated by white arrows. Each line profile has been normalized by the maximum value.}
\label{fig:label-5}
\end{figure}

\section{\label{sec:level1}VOLTAGE- AND POWER- DEPENDENCE OF EMISSION PROCESSES}

\subsection{\label{sec:level1}Lower laser power}
\begin{figure*}[t!]
\begin{center}
\includegraphics[scale=0.38]{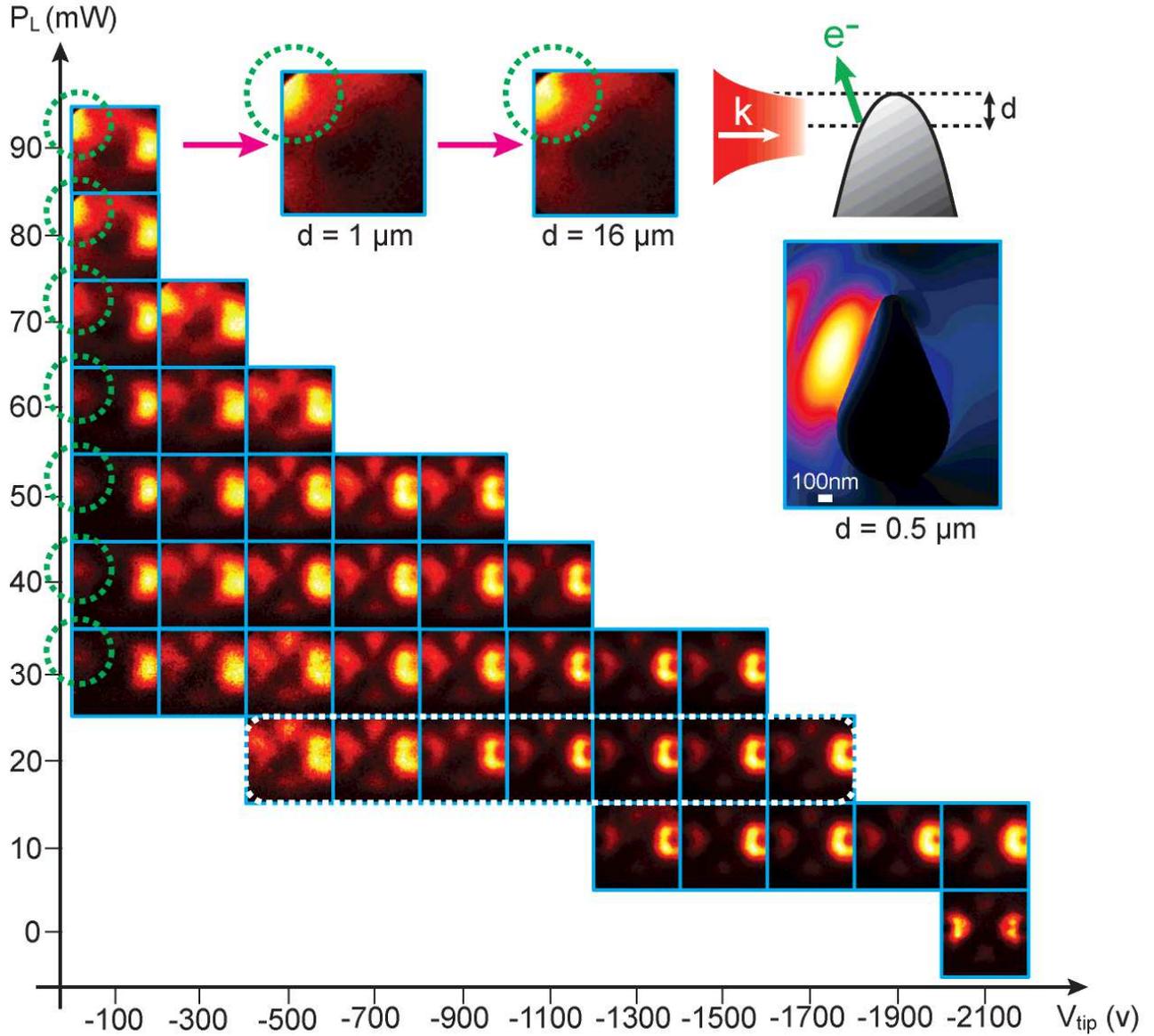}
\end{center}
\vskip -\lastskip \vskip -3pt
\caption{\label{fig:epsart}
(color online). Laser power and tip voltage dependence of the electron emission patterns at [$\varphi = 0^\circ$, $\theta_{P} = 0^\circ$]. On the vertical axis laser power, on the horizontal axis tip voltage are plotted. The inset shows the electron emission patterns where the laser beam is displaced from the tip apex downwards by a distance of 1 $\mu$m and 16 $\mu$m, taken at the 90 mW laser power. The definition of distance d is also shown on the right-hand side of the inset. The time-averaged field distribution around the tip when the laser beam is displaced by a distance of 0.5 $\mu$m is also shown in the inset, where the longer model tip shown in Fig. 2(e) was used. The green dashed circles highlight the left-side electron-emission sites (see section IV B).}
\label{fig:label-5}
\end{figure*}

In this section, we futher discuss the details of electron emission processes and their dependence on laser power and tip voltage by investigating electron emission patterns and Fowler-Nordheim plots, and by simulating electron energy distributions. Figure 10 shows the dependence of LFEM images at [$\varphi = 0^\circ$, $\theta_{P} = 0^\circ$] on the average laser power $P_{L}$ and the tip voltage $V_{tip}$ applied to the tip. Throughout the whole seriese of pictures, the intensity of each image was normalized by the maximum intensity. Normally, total yields decrease as either laser power or tip voltage decrease. As can be seen, the left-right asymmetry is present in all images below a laser power of 60 mW except when $P_{L} =$ 0 mW (FEM). Nevertheless, the images show a trend: the outlines of emission facets become diffuse in the lower tip voltage region, see e.g. the 20 mW row surrounded by a white dashed line. In these experiments, electron emission is considered to be a concerted action of photoemission and field emission. In the case of field emission, the emitted electrons strongly feel the work function corrugation on the nanometer scale, which generates a sharp contrasts at the border of each emission facet. In the case of photoemission, the excited electrons encounter a much narrower surface barrier and appear thus to be less sensitive to the work function corrugation, hence showing a smeared contour of the emission sites. At lower tip voltage, multiphoton processes will be enhanced since field emission is suppressed. Eventually, photoemission from 3PPE or 4PPE will contribute, with energies above the vacuum level. Therefore, the outline of each emission facet becomes diffuse in the lower tip voltage region. This is confirmed by simulations of energy distribution curves and emission patterns.

\begin{figure}[bht]
\begin{center}
\includegraphics[scale=0.24]{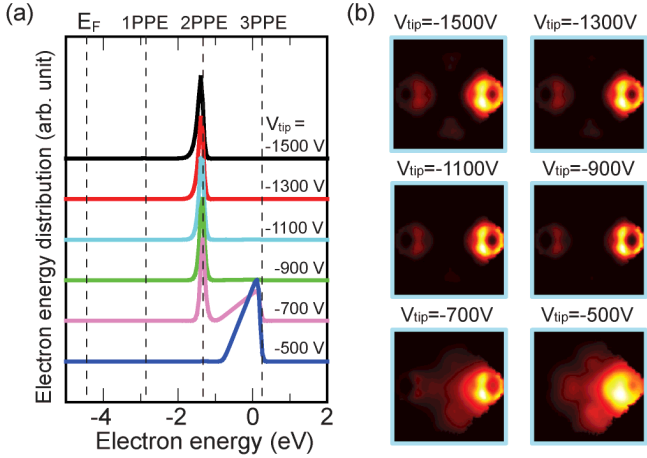}
\end{center}
\vskip -\lastskip \vskip -3pt
\caption{\label{fig:epsart}
(color online). Simulated electron energy distribution curves (a). The spectrum at the top of (a) shows the simulated electron energy distribution of field-emitted electrons with the parameters representing the point where maximum laser intensity can be observed for $\varphi = 0^\circ$, $\theta_{P} = 0^\circ$. The parameters are the following: the work function is 4.45 $eV$, a DC field of 1.32 V/nm corresponds to a tip voltage of -1500 V and $S_{1}$ is $1.6 \cdot 10^{-6}$ for 20 mW laser power. The energy distribution for lower DC fields but the same laser power are also shown in (a). The vacuum level $E_{vac}$ is defined as 0 eV, and the Fermi level $E_{F}$ is 4.45 eV below $E_{vac}$. The energies corresponding to one-, two- and three-photon excitations from $E_{F}$ (1PPE, 2PPE and 3PPE) are also indicated by vertical dashed lines. The simulated LFEM images at corresponding tip voltage and laser power are shown in (b). }
\label{fig:label-5}
\end{figure}

We have simulated the energy distribution of field-emitted electrons with the parameters representing the emission site where maximum intensity can be seen in the simulated image in Fig. 5 at [$\varphi = 0^\circ$, $\theta_{P} = 0^\circ$]: the work function at this point is 4.45 eV, the DC field is 1.32 V/nm and $S_{1}$ is $1.6 \cdot 10^{-6}$, which should correspond to the conditions where the most intense point can be seen in the image at 20 mW laser power and -1500 tip voltage in Fig. 10. Figure 11(a) shows the corresponding simulated energy spectrum at the top, and underneath, spectra for various lower tip voltages. We find that field emission from two-photon processes is strongly dominant at the higher DC fields. For lower DC voltage, the calculated energy distributions clearly show that field-emission from the 2PPE line is suppressed and photoemission from the 3PPE line is enhanced. The simulated LFEM images at corresponding tip voltage are shown in Fig. 11(b). As the DC voltages decrease, the outlines of emission sites become diffuse due to the fact that photoemission processes become dominant. This is in line with the experimental observations described above.

\begin{figure}[t]
\begin{center}
\includegraphics[scale=0.26]{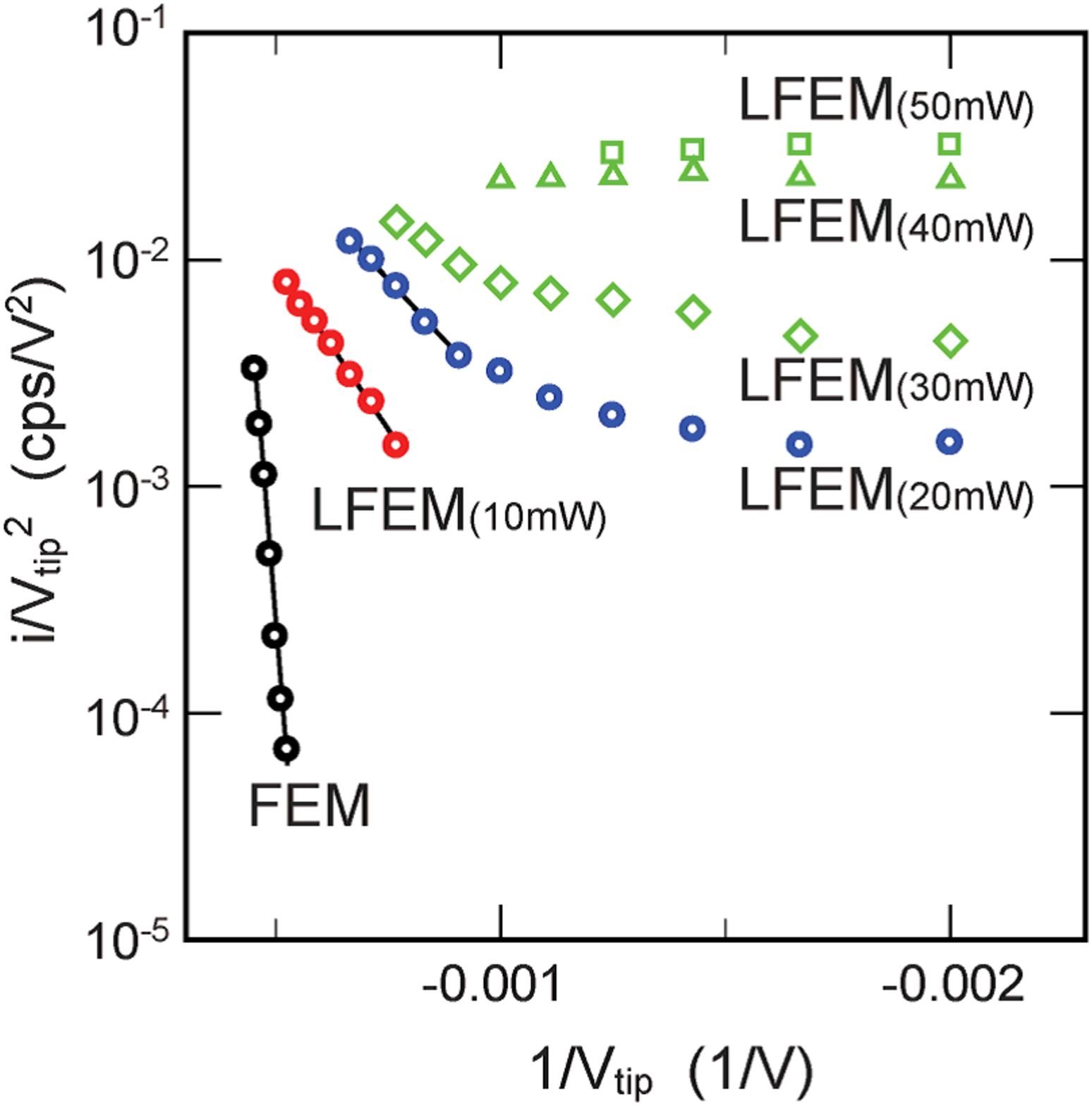}
\end{center}
\vskip -\lastskip \vskip -3pt
\caption{\label{fig:epsart}
(color online). Fowler-Nordheim (FN) plots for FEM and LFEM ($P_{L} =$ 10 mW, 20 mW, 30 mW, 40 mW, 50 mW). The vertical axis is signal $i$ over the $V_{tip}^{2}$ on a logarithmic scale. The signal $i$ is the total yield of electron emission from the right hand (310) type facet of each image in Fig. 10. The horisontal axis is $1/V_{tip}$.}
\label{fig:label-5}
\end{figure}

Experimental Fowler-Nordheim (FN) plots give further suppports for the suggested emission mechanism. Fig. 12 shows FN plots of FEM and LFEM, where the electron count rate $i$ devided by $V_{tip}^{2}$ is displayed on a log scale versus the inverse of $V_{tip}$. The count rate was taken by integrating the electron emission from the right-hand (310) type facet in Fig. 10. According to the FN theory, the linearity of such plots indicates that the electrons are emitted through field emission processes~\cite{gomer93}. The FN plots of FEM data in Fig. 12 clearly show linear behavior, and linearity can be seen also for LFEM at 10 mW and 20 mW, which are shown together with approximated exponential functions (black solid lines). The FN plots for 20 mW show non-linear behavior at low bias voltage. This indicates that photoemission processes become dominant at low voltage as discussed above. Similar behavior is also observed in the case of higher laser power, 30 mW, 40 mW and 50 mW.

The slope of the straight sections is proportional to $\Phi^{3/2}$~\cite{gomer93}. From this fact, we can estimate the effective barrier height $\Phi_{LFEM(10mW)}$ and $\Phi_{LFEM(20mW)}$ at $P_{L} =$ 10 mW and 20 mW, respectively, which an emitted electron feels in the case of LFEM. First, the barrier height ratios were derived from the proportionality constants: $\Phi_{LFEM(10mW)}$/$\Phi_{FEM} =$ 0.24 and $\Phi_{LFEM(20mW)}$/$\Phi_{FEM} =$ 0.2. Taking the work function of (310) type facets of 4.35 eV for $\Phi_{FEM}$, thus $\Phi_{LFEM(10mW)} =$ 1.05 eV and $\Phi_{LFEM(20mW)} =$ 0.85 eV were obtained. The energy difference between $\Phi_{LFEM}$ and $\Phi_{FEM}$ should corresponds to the energy of emitted electrons measured from the Fermi level in LFEM. Here we obtain $\Phi_{LFEM(10mW)}$ - $\Phi_{FEM} =$ 3.3 eV and $\Phi_{LFEM(20mW)}$ - $\Phi_{FEM} =$ 3.5 eV, which is close to the electron energy after two-photon excitation, i.e. 3.1 eV. These values corroborate the two-photon photo-field emission processes, which is consistent with the simulated electron energy distributions for higher voltage in Fig. 11(a).

\subsection{\label{sec:level1}Higher laser power}
For this discussion, we would like to point out the electron emission from the shank side of the tip in the higher laser power range. As in the previous sections, the electron emission from the tip apex is dominant in the emission pattern at low laser power because of the local field enhancement at the tip apex even though the surface area exposed to the laser beam is much larger for the shank than for the apex. At the higher laser power, however, the electron emission from the shank side becomes noticiable because of the nonlinear dependence of the electron emission intensities on the laser power.

In the column of $V_{tip} = -100$V of Fig. 10, the left-side electron-emission features highlighted by green dashed circles becomes suddenly very strong for laser powers exceeding 70 mW. At 90 mW, the intensity of left-side electron emission is comparable to that on the right-side. It remains even when the position of the tip in the beam waist is varied. The insets of Fig. 10 show electron emission patterns at 90 mW laser power where the laser beam is displaced from the tip apex downwards by distance of 1 $\mu$m and 16 $\mu$m. In the two images, right-side emission sites disappear, but the left-side emission remains, indicating that it originates from the shank side of the tip. Such an electron emission should be dominated by photoemission over the surface barrier because DC fields on the shank side are significantly weak with respect to the tip apex. Since the laser pulses arrive at an angle of 20$^\circ$ off the horizontal line in both LFEM images, the position of the electron emission from the shank is also deviated from the horizontal line. In the inset, we also show the time-averaged field distribution around the tip when the laser beam is displaced downwards from the tip apex by a distance of 0.5 $\mu$m: the longer model tip shown in Fig. 2(e) was used. The maximum field can be observed at the side exposed to laser, which is consistent with our observations.

\section{\label{sec:level1}Conclusions}
We have observed laser-induced modulations of field emission intensity distributions resulting in strong asymmetries, which originate from the laser-induced local fields on the tip apex. By varying the laser polarization and the laser incidence direction relative to both azimuthal and polar orientation of the tip apex, we have demonstrated the realization of an ultrafast pulsed field-emission source with convenient control of nanometer sized emission sites. These experimental observations are quantitatively reproduced by using simulated local fields for the photo-field emission model. We discussed the emission processes further and found field-emission after two photon photo-excitation to be the dominant process in laser-induced field emission. From experimental data and simulations, the dependence of the emission processes on laser power and tip voltage could be understood. 

This type of electron source is potentially useful for many applications such as time-resolved electron microscopy, spatio-temporal spectroscopy, near-field imaging techniques, surface-enhanced Raman spectroscopy, or coherent chemical reaction control. Maybe the most interesting applications will arise when two laser pulses with different polarizations or paths are used for the emission of two independent electron beams from two different sites on the tip, spaced only a few tens of nanometers apart, and with an adjustable time delay between the two electron pulses. Since field emission electron sources produce highly coherent electron beams due to their inherently small source size, comparable to the finite spatial extent of electron wave packets inside the source~\cite{oshima02,cho04}, we could expect two spatially and temporally coherent electron beams to be available within the coherence time. This should create new opportunities for addressing fundamental questions in quantum mechanics such as anticorrelation of electron waves in vacuum \cite{klesel02}, or for new directions in electron holography \cite{tonomura87}.

\begin{acknowledgments}We acknowledge many useful discussions with Prof. H. W. Fink, Dr. C. Escher, Dr. T. Ishikawa, and Dr. K. Kamide. This work was supported in part by the Japan Society for the Promotion of Science (JSPS), and the Swiss National Science Foundation (SNSF).
\end{acknowledgments}



\begin{thebibliography}{99}
\bibitem{nagaoka98}K. Nagaoka, T. Yamashita, S. Uchiyama, M. Yamada, H. Fujii, and C. Oshima, Nature (London). {\bf 396}, 557 (1998).

\bibitem{oshima02}C. Oshima, K. Mastuda, T. Kona, Y. Mogami, M. Komaki, Y. Murata, T. Yamashita, T. Kuzumaki, and Y. Horiike, Phys. Rev. Lett. {\bf 88}, 038301 (2002).

\bibitem{cho04}B. Cho, T. Ichimura, R. Shimizu, and C. Oshima, Phys. Rev. Lett. {\bf 92}, 246103 (2004).

\bibitem{gomer93}R. Gomer, "Field Emission and Field Ionization", (American Institute of Physics, New York, 1993).

\bibitem{fursey03}G. Fursey, "Field Emission in Vacuum Microelectronics", (Kluwer Academic / Plenum Publishers, New York, 2003).

\bibitem{fink90}H. -W. Fink, W. Stocker, and H. Schmid, Phys. Rev. Lett. {\bf 65}, 1204 (1990).

\bibitem{fink86}H. W. Fink, IBM. J. Res. Develop. {\bf 30}, 460 (1986).

\bibitem{fu01}T-Y. Fu, L-C. Cheng, C. -H. Nien, and T. T. Tsong, Phys. Rev. B {\bf 64}, 113401 (2001).




\bibitem{hommelhoff06a}P. Hommelhoff, Y. Sortais, A. Aghajani-Talesh, and M. A. Kasevich, Phys. Rev. Lett. {\bf 96}, 077401 (2006).

\bibitem{hommelhoff06b}P. Hommelhoff, C. Kealhofer, and M. A. Kasevich, Phys. Rev. Lett. {\bf 97}, 247402 (2006).

\bibitem{ropers07}C. Ropers, D. R. Solli, C. P. Schulz, C. Lienau, and T. Elsaesser, Phys. Rev. Lett. {\bf 98}, 043907 (2007).

\bibitem{barwick07}B. Barwick, C. Corder, J. Strohaber, N. Chandler-Smith, C. Uiterwaal, and H. Batelaan, New Journal of Physics {\bf 9}, 142 (2007).

\bibitem{wu08}L. Wu, and L. K. Ang, Phys. Rev. B {\bf 78}, 224112 (2008).

\bibitem{aeschlimann07}M. Aeschlimann {\it et al.}, Nature {\bf 446}, 301 (2007).
\bibitem{novotny97}L. Novotny, R. X. Bian, and X. S. Xie, Phys. Rev. Lett. {\bf 79}, 645 (1997).

\bibitem{novotny02}L. Novotny, J. Am. Ceram. Soc. {\bf 85}, 1057 (2002).

\bibitem{hecht05}B. Hecht, L. Novotny, "Principles of Nano-Optics", (Cambridge University Press, Cambridge, 2005).



\bibitem{martin01}Y. C. Martin, H. F. Hamann, and H. K. Wickramasinghe, J. Appl. Phys. {\bf 89}, 5774 (2001).




\bibitem{gao87}Y. Gao, and R. Reifenberger, Phys. Rev. B {\bf 35}, 4284 (1987).


\bibitem{yanagisawa09}H. Yanagisawa, C. Hafner, P. Don\'{a}, M. Kl\"{o}ckner, D. Leuenberger, T. Greber, M. Hengsberger, and J. Osterwalder, Phys. Rev. Lett. {\bf 103}, 257603 (2009).



\bibitem{firester77}A. H. Firester, M. H. Heller, and P. Sheng, Appl. Opt. {\bf 16}, 1971 (1977).

\bibitem{fowles89}G. R. Fowles, "Introduction to Modern Optics" (Dover Publications, New York, ed. 2, 1989).

\bibitem{sato80}M. Sato, Phys. Rev. Lett. {\bf 45}, 1856 (1980).

\bibitem{greber97}T. Greber, O. Raetzo, T. J. Kreutz, P. Schwaller, W. Deichmann, E. Wetli, and J. Osterwalder, Rev. Sci. Instrum. {\bf 68} 4549 (1997).

\bibitem{christian90}C. Hafner, "The Generalized Multipole Technique for Computational Electromagnetics" (Artech House Books, Boston, 1990).

\bibitem{christian99}C. Hafner, "Post-modern Electromagnetics: Using Intelligent MaXwell Solvers" (John Wiley \& Sons, Chichester, 1999).

\bibitem{mmp}http://alphard.ethz.ch/.


\bibitem{max1}http://MaX-1.ethz.ch

\bibitem{christian98}C. Hafner, "MaX-1: A Visual Electromagnetics Platform for PCs" (John Wiley \& Sons, Chichester, 1998).

\bibitem{openmax1}http://OpenMaX.ethz.ch





\bibitem{tungstenepsilon}CRC Handbook of Chemistry and Physics, edited by D. R. Lide (CRC Press, Boca Raton, FL, 2009), 90th ed.



\bibitem{michaelson77}H. B. Michaelson, J. Appl. Phys. {\bf 48}, 4729 (1977).



\bibitem{goldepsilon}P. B. Johnson, and R. W. Christy, Phys. Rev. B {\bf 6}, 4370 (1972).

\bibitem{zenneck}F. Yang, J. R. Sambles, and G. W. Bradberry, Phys. Rev. B {\bf 44}, 5855 (1991).



\bibitem{murphy56}E. L. Murphy, and R. H. Good Jr., Phys. Rev. {\bf 102}, 1464 (1956).

\bibitem{young59}R. D. Young, Phys. Rev. {\bf 113}, 110 (1959).



\bibitem{lisowski04}M. Lisowski {\it et al.}, Appl. Phys. A {\bf 78}, 165 (2004).

\bibitem{merschdorf04}M. Merschdorf, C. Kennerknecht, and W. Pfeiffer, Phys. Rev. B {\bf 70}, 193401 (2004).




\bibitem{Mendenhall34}C. E. Mendenhall, and C. F. DeVoe, Phys. Rev. {\bf 51}, 346 (1937).

\bibitem{hufner95}S. Hufner, Photoelectron Spectroscopy: Principles and Applications (Springer-Werlag, Berlin, 1995).

\bibitem{klesel02}H. Klesel, A, Renz, and F. Hasselbach, Nature {\bf 418}, 392 (2002).

\bibitem{tonomura87}A. Tonomura, Rev. Mod. Phys. {\bf 59}, 639 (1987).

\end{thebibliography}
\end{document}